\def\lb{\hfil\break}
\def\pmb#1{\setbox0=\hbox{#1}%
   \kern-.025em\copy0\kern-\wd0
   \kern.05em\copy0\kern-\wd0
   \kern-0.025em\raise.0433em\box0}
\def\gta{\mathrel{{\lower 3pt\hbox{$\mathchar"218$}}\hskip-8pt
   \raise 2pt\hbox{$\mathchar"13E$}}}
\def\lta{\mathrel{{\lower 3pt\hbox{$\mathchar"218$}}\hskip-8pt
   \raise 2pt\hbox{$\mathchar"13C$}}}
\def\half{{\scriptstyle{1\over2}}}
\def\dagg{\phantom{\dagger}}            
\def\subboldc{\pmb{$\scriptstyle c$}}   
\def\boldd{\pmb{d}}                     
\def\boldell{\pmb{$\ell$}}              
\def\subboldell{\pmb{$\scriptstyle\ell$}}
\def\boldT{\pmb{$T$}}                   
\def\bolddelta{\pmb{$\delta$}}
\def\subbolddelta{\pmb{$\scriptstyle\delta$}}
\def\boldeta{\pmb{$\eta$}}
\def\subboldeta{\pmb{$\scriptstyle\eta$}}
\def\boldxi{\pmb{$\xi$}}
\def\vacuum{|\pmb{\it O}\thinspace\rangle}
\def\up{\uparrow}
\def\dn{\downarrow}
\def\ud{\uparrow\downarrow}
\def\today{\number\day\space\ifcase\month\or
  January\or February\or March\or April\or May\or June\or
  July\or August\or September\or October\or November\or December\fi
 \space\number\year}
\documentstyle[twocolumn,aps]{revtex}
\font\tenrm=cmr8
\begin{document}
\newcommand{\Vv }{{\raisebox{-1.2pt}{\makebox(0,0){$o$}}}}
\newcommand{\Zz }{{\raisebox{-1.2pt}{\makebox(0,0){$\mbox{\tiny o}$}}}}
\newcommand{\Xx }{{\special{em:moveto}}}
\newcommand{\Yy }{{\special{em:lineto}}}
\newcommand{\Ww }{{\usebox{\plotpoint}}}
\title{\begin{minipage}{6.5in}
 {\large \ \\ \ \\ \ \\  \ \\
\centerline{STRIPES, NON-FERMI-LIQUID BEHAVIOR, AND TWO-COMPONENT} 
\centerline{TRANSPORT IN THE HIGH-$T_c$ CUPRATES} }
{\normalsize  \ \\ 
\centerline{J. ASHKENAZI}
\centerline{Physics Department, University of Miami, P.O. Box 248046,
Coral Gables, FL 33124, U.S.A.} \\ \ \\ \ } 
\end{minipage}
}
\author{\ 
\begin{minipage}{5.125in}
\marginparwidth 0.625in 
\small
\baselineskip 9pt
{\bf Abstract}---Non-Fermi-liquid features of the high-$T_c$ cuprates,
and specifically the systematic behavior of the resistivity, Hall
constant, and thermoelectric power, are shown to result from an
electronic structure based on ``large-$U$'' and ``small-$U$'' orbitals,
and the resulting striped structure. 
\end{minipage}
}
\maketitle
\setlength{\unitlength}{1in}
\makeatletter
\global\@specialpagefalse
\def\@oddhead{\footnotesize \it \hfill \ Journal of Physics and 
Chemistry of Solids}
\makeatother
\baselineskip 12pt
\normalsize \rm
%

The electronic structure of the cuprates is studied in term of
``large-$U$'' and ``small-$U$'' orbitals [1]. The ``slave-fermion''
method is applied, where the creation operator of a large-$U$ electron
in site $i$ and spin $\sigma$ is expressed as $e_i^{\dagger}
s_{i,-\sigma}^{\dagg}$, if it is in the ``upper-Hubbard-band'', and as
$\sigma s_{i\sigma}^{\dagger} h_i^{\dagg}$, if it is in a
Zhang-Rice-type ``lower-Hubbard-band''; $e_i^{\dagg}$ and $h_i^{\dagg}$
are (``excession'' and ``holon'') fermion operators, and
$s_{i\sigma}^{\dagg}$ are (``spinon'') boson operators, and the
constraint $e_i^{\dagger} e_i^{\dagg} + h_i^{\dagger} h_i^{\dagg} +
\sum_{\sigma} s_{i\sigma}^{\dagger} s_{i\sigma}^{\dagg} = 1$ should be
satisfied. 

Here this constraint is imposed on the average within an auxiliary
Hilbert space, and physical observables are calculated by taking
appropriate combinations of Green's functions of this space. The Green's
functions are determined by the Hamiltonian which obeys the constraint
rigorously. Two-particle spinon-holon Green's functions are decoupled
only where the ``spin-charge separation'' approximation holds. 


The Bogoliubov transformation is applied to diagonalize the spinons,
yielding creation operators $\zeta_{\sigma}^{\dagger} ({\bf k})$, and
``bare'' spinon energies $\epsilon^{\zeta} ({\bf k})$ with a V-shape
zero minimum at ${\bf k}={\bf k}_0$, where $2{\bf k}_0 = ( {\pi \over
{\rm a}} , {\pi \over {\rm a}} )$. Bose condensation results in
antiferromagnetism (AF). 

Numerical calculations in a lightly doped AF plane, supported by
neutron-scattering measurements in some cuprates [2], indicate the
existence of a frustrated striped structure where narrow charged stripes
form antiphase domain walls separating wider AF stripes. Growing
experimental evidence supports the assumption that such a structure
exists, at least dynamically, in all the superconducting cuprates. 

Since the spin-charge separation approximation is valid in
one-dimension, it should apply for holons within the charged stripes,
and they are referred to as ``stripons'', created by
$p^{\dagger}_{\mu}({\bf k})$, and of bare energies
$\epsilon^p_{\mu}({\bf k})$. Since one expects finite stripe segments,
frustrations, and defects, which are fatal for itinerancy in
one-dimension, we assume a starting point of localized stripon states. 

The holon-spinon and excession-spinon pair states for which spin-charge
separation does not apply hybridize with the small-$U$ electron states
forming, within the auxiliary space, states of ``Quasi-electrons''
(QE's), created by $q_{\iota\sigma}^{\dagger}({\bf k})$. Their bare
energies $\epsilon^q_{\iota} ({\bf k})$ form quasi-continuous ranges of
bands crossing the Fermi level ($E_{_{\rm F}}$) over ranges of the
Brillouin zone (BZ). 


These quasiparticles are coupled due to hopping and hybridization terms, 
and the coupling between them can be expressed through a Hamiltonian term: 
\begin{eqnarray}
{\cal H}^{\prime} &=& {1 \over \sqrt{N}} \sum_{\iota\mu\lambda\sigma}
\sum_{{\bf k}, {\bf k}^{\prime}} \Big\{\sigma
\epsilon^{qp}_{\iota\mu\lambda\sigma}({\bf k}^{\prime}, {\bf k})
q_{\iota\sigma}^{\dagger}({\bf k}) p_{\mu}^{\dagg}({\bf k}^{\prime})
\nonumber \\ &\ &\times\big[ \cosh{(\xi_{\lambda\sigma,({\bf k} - {\bf
k}^{\prime})})} \zeta_{\lambda\sigma}^{\dagg}({\bf k} - {\bf
k}^{\prime}) \nonumber \\ &\ &+ \sinh{(\xi_{\lambda\sigma,({\bf k} -
{\bf k}^{\prime})})} \zeta_{\lambda,-\sigma}^{\dagger}({\bf k}^{\prime}
- {\bf k}) \big] + h.c. \Big\}.
\end{eqnarray} 
${\cal H}^{\prime}$ introduces a vertex between QE, stripon and spinon
propagators. Vertex corrections are negligible by a generalized Migdal
theorem, since the obtained stripon bandwidth is much smaller than the
QE and spinon bandwidths. Thus a second-order perturbation expansion in
${\cal H}^{\prime}$ is applicable. The QE, spinon, and stripon
scattering rates 
$\Gamma^q_{\iota}({\bf k}, \omega)$, $\Gamma^{\zeta}_{\lambda}({\bf k},
\omega)$, and $\Gamma^p_{\mu}({\bf k}, \omega)$, 
are then calculated, and for sufficiently doped cuprates one gets a
self-consistent solution of the following features: 

Spinons: Their spectral function $A^{\zeta}({\bf k},
\omega)\propto\omega$ for small $\omega$, and thus $A^{\zeta}({\bf k},
\omega) b_{_T}(\omega)\propto T$ for $\omega\ll T$, where
$b_{_T}(\omega)$ is the Bose distribution function. 

Stripons: Their localized states are renormalized to polaron-like states
very close to $E_{_{\rm F}}$, with some hopping through QE-spinon
states. One gets $\Gamma^p({\bf k}, \omega) \propto A \omega^2 + B
\omega T + CT^2$, and a two-dimensional itinerant behavior at low
temperatures, with a bandwidth of $\sim$$0.02\;$eV. 

Quasi-electrons: An approximate expression for their scattering rates is
given by $\Gamma^q({\bf k}, \omega) \propto \omega[b_{_T}(\omega) +
\half]$, becoming $\Gamma^q({\bf k}, \omega)\propto T$ in the limit
$T\gg |\omega|$, and $\Gamma^q({\bf k}, \omega)\propto\half |\omega|$ in
the limit $T\ll |\omega|$, as in marginal-Fermi-liquid phenomenology. 

Lattice effects (``svivons''): The charged stripes are characterized by
an LTT-like structure [3]. Thus, spinon excitations due to ${\cal
H}^{\prime}$ are followed by phonon excitations, and stripons have
polaron-like lattice features. A spinon propagator linked to a vertex is
thus ``dressed'' by phonon propagators. We refer to such a
phonon-dressed spinon as a svivon. 



The dc current is expressed as a sum ${\bf j} = {\bf j}^q + {\bf j}^p$
of QE and stripon currents. Since stripons hop only
via QE states, one gets that ${\bf j}^p \cong \alpha {\bf j}^q$, where
$\alpha$ is $T$-independent. Consequently, an electric
field is accompanied by gradients {\bf \pmb{$\nabla$}\/}$\mu^q$ and {\bf
\pmb{$\nabla$}\/}$\mu^p$ of the QE and stripon chemical potentials,
satisfying $N^q${\bf \pmb{$\nabla$}\/}$\mu^q + N^p${\bf
\pmb{$\nabla$}\/}$\mu^p = 0$, where $N^q$ and $N^p$ are the
contributions of QE and stripon states to the electrons density of
states at $E_{_{\rm F}}$. 

By using the Kubo formula we derive expressions for the dc conductivity
and Hall constant, in terms of Green's functions. These expressions
include diagonal and non-diagonal conductivity QE and stripon terms
$\sigma_{xx}^{qq}$, $\sigma_{xx}^{pp}$, $\sigma_{xy}^{qqq}$,
$\sigma_{xy}^{ppp}$, and mixed terms $\sigma_{xy}^{qqpp}$. The currents
are expressed as: $j_x^q=\sigma_{xx}^{qq} {\cal E}_x^q$,
$j_x^p=\sigma_{xx}^{pp} {\cal E}_x^p$, where {\bf \pmb{${\cal
E}$}\/}$^q={\bf E} + ${\bf \pmb{$\nabla$}\/}$\mu^q/{\rm e}$, {\bf
\pmb{${\cal E}$}\/}$^p={\bf E} + ${\bf \pmb{$\nabla$}\/}$\mu^p/{\rm e}$,
and ${\bf E}$ is the electric field. By expressing ${\bf E}=(N^q ${\bf
\pmb{${\cal E}$}\/}$^q + N^p $ {\bf \pmb{${\cal E}$}\/}$^p)/(N^q+N^p)$,
and $j^q_x + j^p_x = j_x = E_x/\rho_x$, one gets that the resistivity
can be expressed as: 
\begin{equation}
\rho_x = {1 \over (N^q+N^p) (1+\alpha)} \Big( {N^q \over
\sigma_{xx}^{qq}} + {\alpha N^p \over \sigma_{xx}^{pp}}\Big).  
\end{equation}
Similarly, the Hall constant $R_{_{\rm H}} = E_y/j_xH$ can be 
expressed as $R_{_{\rm H}} = \rho_x / \cot{\theta_{_{\rm H}}}$, 
where:
\begin{equation}
{(1+\alpha) \over \cot{\theta_{_{\rm H}}} } = \Big[{\sigma_{xy}^{qqq} +
\sigma_{xy}^{qqpp} \over \sigma_{xx}^{qq}} + {\alpha(\sigma_{xy}^{ppp} +
\sigma_{xy}^{qqpp}) \over \sigma_{xx}^{pp}} \Big].
\end{equation}

To get the temperature dependencies of $\rho$ and $\cot{\theta_{_{\rm
H}}}$ we use those derived for $\Gamma^q$ and $\Gamma^p$ (to which
temperature-independent impurity scattering terms are added). Thus one
can parametrize: $\sigma_{xx}^{qq}\propto (D+CT)^{-1}$,
$\sigma_{xx}^{pp}\propto (A+BT^2)^{-1}$, $\sigma_{xy}^{qqq}\propto
(D+CT)^{-2}$, $\sigma_{xy}^{ppp}\propto (A+BT^2)^{-2}$,
$\sigma_{xy}^{qqpp}\propto [(D+CT)(A+BT^2)]^{-1}$, and express: 
\begin{eqnarray}
\rho &\cong& {(D+CT+A+BT^2) \over N}, \\ \cot{\theta_{_{\rm H}}}
&\cong& \Big({Z \over D+CT} + {1 \over A+BT^2} \Big)^{-1}. 
\end{eqnarray} 

These expressions reproduce the systematic behavior of $\rho$ and
$\cot{\theta_{_{\rm H}}}$ in different cuprates, except for the effect
of the pseudogap, not included in this parametrization. Results
corresponding to data in YBa$_2$Cu$_{3-x}$Zn$_x$O$_7$ [4],
Tl$_2$Ba$_2$CuO$_{6+\delta}$ [5], and La$_{2-x}$Sr$_x$CuO$_4$ [6], are
presented in Figs. 1, 2, and 3, respectively. 

\vskip -1.1truecm
\begin{figure}[t]
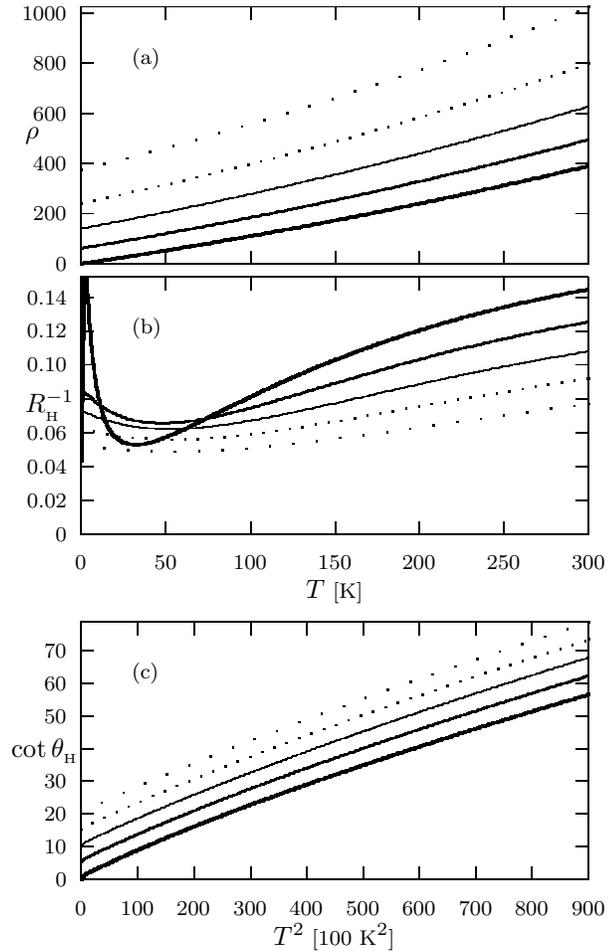


\setlength{\unitlength}{0.240900pt}
\ifx\plotpoint\undefined\newsavebox{\plotpoint}\fi
\sbox{\plotpoint}{\rule[-0.175pt]{0.350pt}{0.350pt}}%



\caption{The transport coefficients, in arbitrary unit, for
A=1,7,13,19,25; B=.001; C=1; D=0,50,100,150,200; N=1,.9,.8,.7,.6; Z=2.
The first value corresponds to the thickest lines.} 
\label{F1}
\end{figure}

The idea of different scattering rates for $\rho$ and
$\cot{\theta_{_{\rm H}}}$ has been first suggested by Anderson [7],
where in his analysis the $T^2$ term is due to spinons. Note however
that in recent ac Hall effect results [8] the energy scale corresponding
to this term is found to be $\sim$$120\;$K, in agreement with that of
our stripons, rather than the much higher spinon energies. 
 
Under a temperature gradient one gets: ${\bf j}^q={\rm e}T^{-1}
\underline{\bf L}^{q(11)} ${\bf \pmb{${\cal E}$}\/}$^q +\underline{\bf
L}^{q(12)} ${\bf \pmb{$\nabla$}\/}$(T^{-1})$, \ ${\bf j}^p={\rm e}T^{-1}
\underline{\bf L}^{p(11)} ${\bf \pmb{${\cal E}$}\/}$^p +\underline{\bf
L}^{p(12)} ${\bf \pmb{$\nabla$}\/}$( T^{-1})$. The thermoelectric power
(TEP) is given by $\underline{\bf S}=[{\bf E} / ${\bf
\pmb{$\nabla$}\/}$T]_{{\bf j} = 0}$. Since ${\bf j}^p\cong\alpha {\bf
j}^q$, the condition ${\bf j}=0$ means ${\bf j}^q\cong{\bf j}^p\cong 0$.
Thus one gets: $\underline{\bf S}=(N^q\underline{\bf S}^q +
N^p\underline{\bf S}^p) / (N^q + N^p)$, where $\underline{\bf S}^q=-
\underline{\bf L}^{q(12)} / {\rm e}T \underline{\bf L}^{q(11)}$, \
$\underline{\bf S}^p=- \underline{\bf L}^{p(12)} / {\rm e}T
\underline{\bf L}^{p(11)}$. 

\vskip -1.1truecm
\begin{figure}[t]
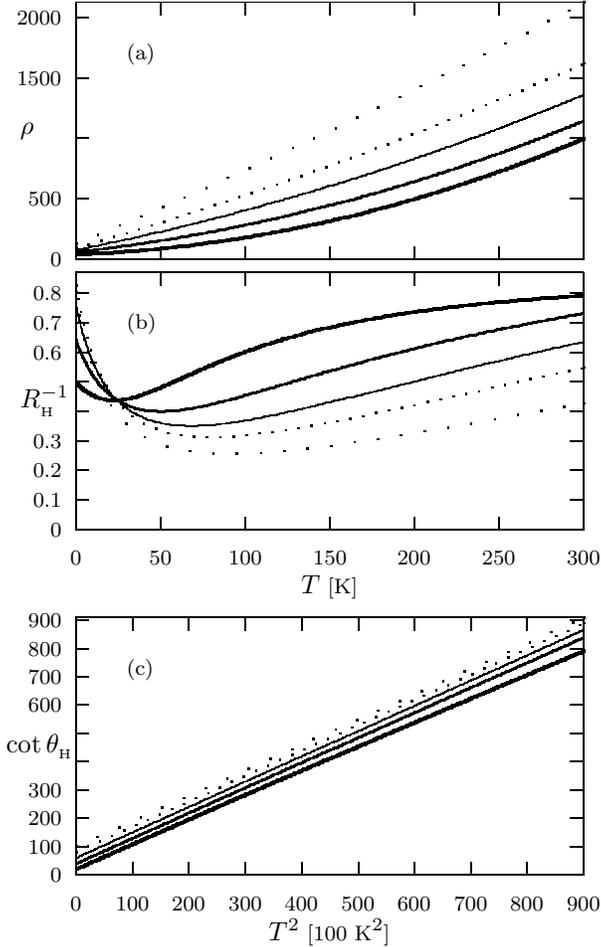


\setlength{\unitlength}{0.240900pt}
\ifx\plotpoint\undefined\newsavebox{\plotpoint}\fi
\sbox{\plotpoint}{\rule[-0.175pt]{0.350pt}{0.350pt}}%



\caption{The transport coefficients, in arbitrary unit, for
A=20,40,60,80,100; B=.01; C=.5,2,5,10,20; D=20,40,80,160,320;
N=1,1.3,1.8,2.5,3.4; Z=.01. The first value corresponds to the thickest
lines.} 
\label{F2}
\end{figure}

One gets $S^q\propto T$, as for electrons in metals, while $S^p$
saturates at $T \simeq 200\; $K to the narrow-band result:
$S^p=(k_{_{\rm B}}/{\rm e})\ln{[(1}-n^p)/n^p]$, where $n^p$ is the
fractional occupation of the stripon band. This is consistent with the
typical behavior of the TEP in the cuprates. It was found [9] that
$S^p=0$ (and thus $n^p=0.5$) for slightly overdoped cuprates. 


Considering large-$U$ and small-$U$ orbitals in the cuprates results in
a striped structure, and three types of quasiparticles: polaron-like
stripons carrying charge, phonon-dressed spinons (svivons) carrying
spin, and quasi-electrons carrying both. Non-Fermi-liquid features of
the cuprates are explained, and specifically the systematic behavior of
the resistivity, Hall constant, and thermoelectric power. 


\vskip -1.1truecm
\begin{figure}[t]
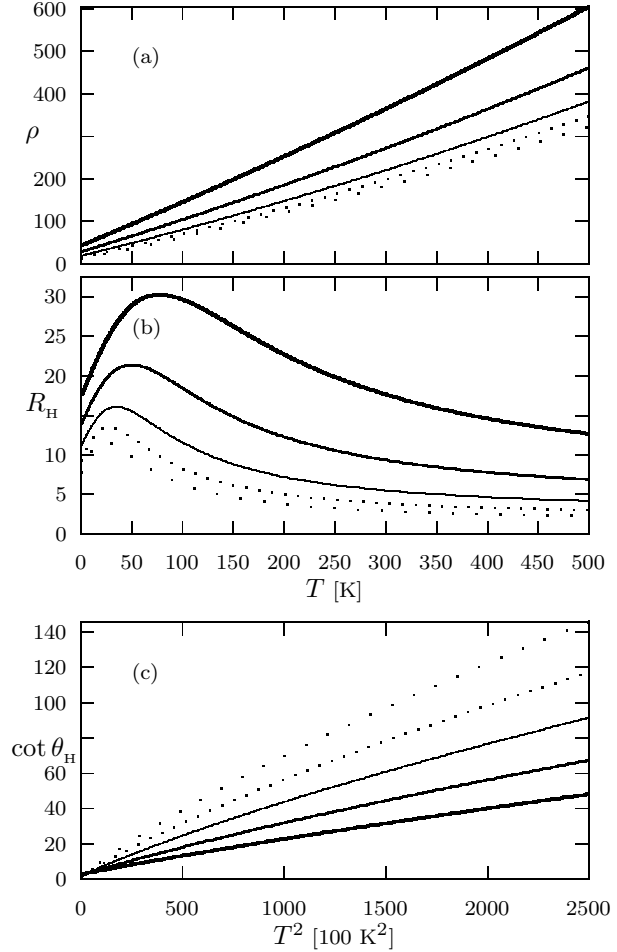


\setlength{\unitlength}{0.240900pt}
\ifx\plotpoint\undefined\newsavebox{\plotpoint}\fi
\sbox{\plotpoint}{\rule[-0.175pt]{0.350pt}{0.350pt}}%



\caption{The transport coefficients, in arbitrary unit, for
A=3,2.4,2,1.7,1.5; B=.00025,.0004,.0006,.0008,.001; C=1,1.1,1.3,1.6,2;
D=40; N=1,1.5,2.2,3,4; Z=3. The first value corresponds to the thickest
lines.} 
\label{F3}
\end{figure}

\centerline{\bf REFERENCES} 
\medskip
\baselineskip 10pt
\footnotesize
\noindent
1. J.~Ashkenazi, {\it J.~Supercond.} {\bf 10} (Aug. 1997). \\ 
2. J.~M.~Tranquada {\it et al.}, {\it Phys.~Rev.~B} {\bf 54}, 7489
(1996). \\ 
3. A.~Bianconi {\it et al.}, {\it Phys.~Rev.~B} {\bf 54}, 12018 (1996).
\\ 
4. T.~R.~Chien, {\it et al.}, {\it Phys.~Rev.~Lett.} {\bf 67}, 2088
(1991). \\ 
5. Y.~Kubo and T.~Manako, {\it Physica C} {\bf 197}, 378 (1992). \\
6. H.~Takagi, {\it et al.}, {\it Phys.~Rev.~Lett.} {\bf 69}, 2975
(1992); \\ \mbox{\ } \ \ H.~Y.~Hwang, {\it et al.}, {\it ibid.} {\bf
72}, 2636 (1994). \\
7. P.~W.~Anderson, {\it Phys.~Rev.~Lett.} {\bf 67}, 2092 (1991). \\ 
8. H.~D~Drew, this issue. \\
9. B.~Fisher, {\it et al.}, {\it J. Supercond.} {\bf 1}, 53 (1988); \\
\mbox{\ } \ \ S.~D.~Obertelli, {\it et al.}, {\it Phys.~Rev.~B} {\bf
46}, 14928 (1992). 

\end{document}